\documentclass[trackchanges]{aastex701}

\begin{document}

\title{Oscillation Period Properties of Sunspots in the 8-10 $\mu$m Infrared Band: A Multi-Dataset Analysis}

\author{Suo Liu}
\affiliation{State Key Laboratory of Solar Activity and Space Weather, National Astronomical Observatories, Chinese Academy of Sciences, Beijing, 100101, China}
\affiliation{School of Astronomy and Space Sciences, University of Chinese Academy of Sciences, Beijing, 100049, China}
\email{lius@nao.cas.cn}

\begin{abstract}
Sunspot oscillations provide a unique probe of magnetic-plasma interactions in the solar atmosphere. This study investigates oscillation period properties of six single sunspot active regions observed in the 8-10 $\mu$m infrared band using the Accurate Infra-red Magnetic-field Solar Telescope (AIMS) at Lenghu Observatory. Light curves were extracted from three regions (umbra, penumbra, quiet Sun) and wavelet analysis was applied to determine dominant and weighted mean periods, comparing four methods: pixel-wise wavelet analysis with mean aggregation, pixel-wise wavelet analysis with median aggregation, spatial mean prior to wavelet analysis, and spatial median prior to wavelet analysis.
The weighted mean period appears more physically meaningful than the dominant period, better representing the multi-mode nature of solar oscillations. A consistent period sequence holds across all six datasets: umbra (U) $<$ penumbra (P) $<$ quiet Sun (Q), with typical values of 260-313 s, 286-374 s, and 294-382 s, respectively. This key pattern is recovered with 100\% consistency by pixel-wise wavelet analysis with mean aggregation, pixel-wise wavelet analysis with median aggregation, and spatial median prior to wavelet analysis. The spatial mean prior to wavelet analysis shows 1/6 full consistency (complete U $<$ P $<$ Q hierarchy), 3/6 partial consistency (U $<$ P and U $<$ Q satisfied, but P $\ge$ Q), and 2/6 inconsistency (U $>$ P and/or U $>$ Q), with the partial cases preserving the critical umbral period being shortest.
Spatial smoothing generally increases measured periods (by up to 21\%), with the umbra showing the highest sensitivity. All peak ratios fall below 0.3 (mean $0.14 \pm 0.03$), confirming the inherent multi-mode nature of solar oscillations. These findings provide important observational constraints for models of magneto-convection and wave propagation in sunspot atmospheres, establishing the 8-10 $\mu$m band as a valuable diagnostic window for solar physics.
\end{abstract}

\section{Introduction}
\label{sec:intro}

Sunspots are the most prominent manifestations of solar magnetic activity, characterized by intense magnetic fields (2000-3000 G in the umbra) that suppress convective energy transport and modify the propagation of magnetohydrodynamic (MHD) waves. Oscillations in sunspots have been a subject of intensive study since their discovery by \citeauthor{1962ApJ...135..474L} (\citeyear{1962ApJ...135..474L}), providing a unique tool for probing the subsurface structure and atmospheric dynamics of active regions through the framework of local helioseismology. These oscillations are believed to be driven by trapped acoustic waves (p-modes) that interact with the strong magnetic field, leading to mode conversion, absorption, and the excitation of magneto-acoustic waves \citep{1994ApJ...437..505C, 1997RSPSA.453..943B, 2000SoPh..192..373B, 2006ApJ...640.1153C}.

\subsection{Regional Variation of Sunspot Oscillations}

The physical properties of sunspot oscillations vary significantly across different regions, reflecting the spatial structure of the magnetic field. The umbra, with its strong vertical magnetic field (typically 2000--3000~G), exhibits oscillations with periods of approximately 5~minutes in the photosphere \citep{1962ApJ...135..474L, 1970ApJ...162..993U, 1984ApJ...277..874L} and shorter periods of approximately 3~minutes, often referred to as ``umbral flashes'', in the chromosphere \citep{1969SoPh....7..351B, 1984ApJ...277..874L, 2000SoPh..192..373B, 2017ApJ...836...18C}. These oscillations are interpreted as upwardly propagating slow magneto-acoustic waves guided along the nearly vertical magnetic field lines.

Recent observations have confirmed that umbral oscillations persist throughout the entire atmospheric column, with the dominant oscillation period decreasing from the photosphere ($\sim$5~min) to the chromosphere ($\sim$3~min) \citep{2017ApJ...836...18C, 2023ApJ...958...10W}. This behavior is a consequence of the increasing acoustic cutoff frequency with height: in the higher, less dense layers of the chromosphere, the atmospheric cutoff frequency rises, thereby permitting shorter-period waves to propagate upward. The observed period decrease thus provides an important diagnostic for atmospheric stratification and magneto-acoustic wave propagation models.

The penumbra, where the magnetic field becomes more inclined (40-70$^\circ$ from the vertical) and weaker (1000-1500 G), shows a mixture of 3-minute and 5-minute oscillations \citep{1999ASPC..184..103C, 2004A&A...424..671K, 2017NewA...51...86Z, 1982Natur.297..485T}. This intermediate behavior reflects the transition from the strongly magnetized umbra to the weakly magnetized region. The penumbral filaments and the Evershed flow add additional complexity to the oscillation patterns in this region. Recent high-resolution observations have revealed that penumbral oscillations are organized into azimuthal wave modes, with running penumbral waves propagating outward from the umbra-penumbra boundary \citep{2017ApJ...836...18C}.

The surrounding quiet Sun is dominated by the well-known 5-minute p-mode oscillations, which are global acoustic modes of the Sun \citep{1962ApJ...135..474L}. However, the power distribution of these oscillations is not uniform, with enhanced power in the intergranular lanes where downflows concentrate acoustic sources \citep{2006ARep...50..588K}. Photospheric observations in the 8-10 $\mu$m band are particularly valuable because they probe the upper photosphere where the magnetic field-plasma coupling is strongest, providing a critical link between deeper and higher atmospheric layers \citep{2017A&A...605A.125S}.

\subsection{Wavelength Dependence and Infrared Advantages}

The observed oscillation properties depend critically on the wavelength of observation, as different spectral bands probe different atmospheric heights. Visible and near-infrared observations (400-1600 nm) have been extensively used to study sunspot oscillations \citep{1991GApFD..62..153D, 2003SoPh..218...85B}. However, the 8-10 $\mu$m mid-infrared band offers several unique advantages:

1. Deeper atmospheric probing: The 8-10 $\mu$m continuum forms in the upper photosphere (approximately 200-400 km above $\tau_{500} = 1$), deeper than most visible and near-infrared lines \citep{2017A&A...605A.125S}. This allows us to probe the layer where magnetic fields and plasma motions are most strongly coupled.

2. Reduced scattering: Infrared observations benefit from decreased Rayleigh scattering ($\propto \lambda^{-4}$), resulting in higher image contrast and better seeing conditions.

3. Enhanced magnetic sensitivity: The mid-infrared continuum is sensitive to temperature fluctuations induced by magneto-convective motions, providing a direct tracer of the magnetic field's influence on plasma dynamics.

4. Unique observational window: Very few systematic studies of sunspot oscillations have been conducted in the 8-10 $\mu$m band, making this work a valuable addition to the multi-wavelength understanding of sunspot physics.

\subsection{Motivation and Objectives}

This study aims to present new observations of sunspot oscillations in the underexplored 8-10 $\mu$m band. The unique dataset from infrared instrument Accurate Infra-red Magnetic-field Solar Telescope (AIMS) at Lenghu Observatory provides an opportunity to:

1. Present the first systematic study of sunspot oscillations in the 8-10 $\mu$m mid-infrared band for six active regions (NOAA AR 13879, 14087, 14191, 14323, 14336, and 14349), establishing baseline period measurements for this wavelength range.

2. Compare four different analysis methods (Pixel Mean: pixel-wise wavelet analysis with mean aggregation; Pixel Median: pixel-wise wavelet analysis with median aggregation; Spatial Mean: spatial mean prior to wavelet analysis; Spatial Median: spatial median prior to wavelet analysis) to assess the robustness of measured periods and identify the most reliable approach.

3. Investigate the spatial structure of oscillations by examining the period hierarchy across umbra, penumbra, and quiet Sun regions, testing the hypothesis that stronger magnetic fields correspond to shorter oscillation periods.

4. Quantify the effect of spatial smoothing on measured periods, providing guidance for comparing observations from instruments with different spatial resolutions.

The remainder of this paper is organized as follows: Section \ref{sec:obser} describes the observational data, region segmentation, and wavelet analysis methods. Section \ref{sec:results} presents the results, including period comparisons across regions, methods, and smoothing scales. Section \ref{sec:disc} discusses the physical implications of our findings, and comparisons with previous multi-wavelength studies. Our discussions and conclusions are summarized in Section \ref{sec:disc} and \ref{sec:concl}, respectively.

\section{Observations and Methods}
\label{sec:obser}

\subsection{Observational Data}
The data used in this study were obtained with the Accurate Infra-red Magnetic-field Solar Telescope (AIMS) at Lenghu Observatory, National Astronomical Observatories, Chinese Academy of Sciences. The AIMS is a 1-meter aperture telescope equipped with a liquid nitrogen-cooled infrared imaging system operating in the 8-10 $\mu$m mid-infrared band. The cryogenic system maintains the detector at low temperatures, effectively reducing thermal background noise and dark current, which is critical for high-precision observations in this wavelength range. The instrument has a field of view of 256 $\times$ 256 pixels with a plate scale of approximately 1.5 arcsec pixel$^{-1}$, corresponding to a total field of view of approximately $384 \times 384$ arcsec$^{2}$ ($6.4 \times 6.4$ arcmin$^{2}$) on the solar surface \citep{2025RAA....25g5020D}.

A total of six active regions were analyzed in this study, spanning observations from 2024 to 2026. Table \ref{tab:datasets} summarizes the basic information for each dataset, including the NOAA active region number, observation date, and duration. Each dataset consists of 1001 time frames covering approximately 36 minutes, with a cadence of about 2.2 seconds. This temporal resolution is sufficient to resolve the 3-6 minute oscillation periods typical of sunspot oscillations.

\begin{table}[htbp]
\centering
\caption{Summary of observational datasets analyzed in this study.}
\label{tab:datasets}
\begin{tabular}{lcccc}
\hline
\textbf{Dataset} & \textbf{NOAA AR} & \textbf{Observation Date} & \textbf{Duration (s)} & \textbf{Cadence (s)} \\
\hline
Dataset 1 & NOAA 13879 & 2024-11-04 & 2187.5 & 2.19 \\
Dataset 2 & NOAA 14087 & 2025-05-18 & 2187.6 & 2.19 \\
Dataset 3 & NOAA 14191 & 2025-08-23 & 2187.6 & 2.19 \\
Dataset 4 & NOAA 14323 & 2025-09-29 & 2191.4 & 2.19 \\
Dataset 5 & NOAA 14336 & 2025-10-29 & 2209.7 & 2.21 \\
Dataset 6 & NOAA 14349 & 2026-01-25 & 2187.7 & 2.19 \\
\hline
\end{tabular}
\end{table}

\subsection{Region Segmentation}
\begin{figure}[htbp]
\centering
\includegraphics[width=0.95\textwidth]{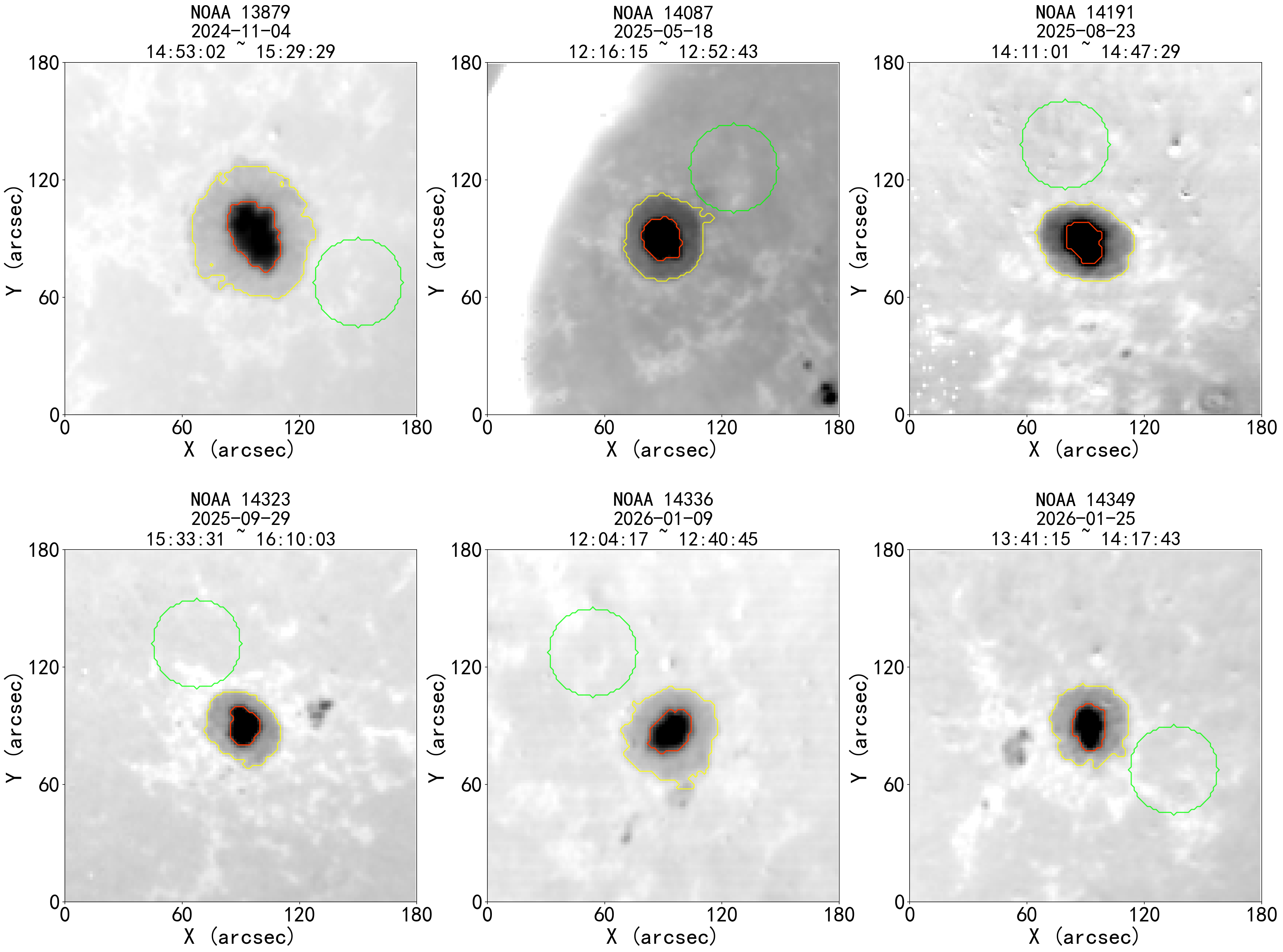}
\caption{Sunspot classification maps for all six datasets. \textbf{Top row:} NOAA AR 13879, 14087, and 14191. \textbf{Bottom row:} NOAA AR 14323, 14336, and 14349. Red contours indicate the umbra, yellow contours indicate the penumbra, and green circles indicate the quiet Sun reference region. The segmentation thresholds were optimized for each dataset to account for varying contrast conditions. All observations were obtained with a cadence of $\sim$2.2 s, covering approximately 36 minutes of continuous monitoring. Each panel shows a cropped region of $120\times120$ pixels (centered on the sunspot), which is approximately $180\times180$ arcsec$^{2}$ given the plate scale of 1.5 arcsec pixel$^{-1}$.}
\label{fig:classification}
\end{figure}

For each sunspot, three distinct regions were identified: umbra, penumbra, and quiet Sun. Figure~\ref{fig:classification} shows the sunspot classification maps for all six datasets. Each panel displays a cropped region of $120\times120$ pixels ($180\times180$ arcsec$^{2}$, $3\times3$ arcmin$^{2}$) centered on the sunspot. The full detector field of view is $256\times256$ pixels ($384\times384$ arcsec$^{2}$, $6.4\times6.4$ arcmin$^{2}$) with a plate scale of 1.5 arcsec pixel$^{-1}$.

The umbra was identified using a region growing algorithm seeded at the darkest pixel within the sunspot. The intensity threshold for umbra was empirically determined for each dataset (ranging from 0.85 to 0.95 in normalized intensity units) to ensure consistent identification across different active regions. Table \ref{tab:segmentation} summarizes the segmentation parameters and pixel statistics for all datasets.

The penumbra was grown outward from the umbra boundary using an iterative dilation process. Pixels with intensity below a penumbral threshold (typically 0.97-0.999 in normalized units) were included in the penumbral region. This approach ensures that the penumbra is contiguous with the umbra and captures the fibril structure typical of penumbral magnetic fields.

The quiet Sun region was defined as a circular area of radius 15 pixels (approximately 22 arcsec) located away from the sunspot, with care taken to avoid including any penumbral or umbral pixels. This region serves as a reference for non-magnetic solar oscillations.

\begin{table}[htbp]
\centering
\caption{Region segmentation parameters and pixel statistics for all six datasets.}
\label{tab:segmentation}
\begin{tabular}{lccccccc}
\hline
\textbf{Dataset} & \textbf{NOAA} & \textbf{Umbra} & \textbf{Penumbra} & \textbf{Umbra} & \textbf{Penumbra} & \textbf{Quiet} & \textbf{Duration} \\
 & \textbf{AR}& \textbf{Threshold} & \textbf{Threshold} & \textbf{Pixels} & \textbf{Pixels} & \textbf{Pixels} & \textbf{(min)} \\
\hline
1 & NOAA 13879 & 0.85 & 0.97 & 309 & 1113 & 709 & 36.5 \\
2 & NOAA 14087 & 0.93 & 0.99 & 153 & 485 & 709 & 36.5 \\
3 & NOAA 14191 & 0.90 & 0.98 & 133 & 559 & 709 & 36.5 \\
4 & NOAA 14323 & 0.92 & 0.999 & 94 & 337 & 709 & 36.5 \\
5 & NOAA 14336 & 0.89 & 0.97 & 64 & 291 & 709 & 36.8 \\
6 & NOAA 14349 & 0.95 & 0.999 & 141 & 440 & 709 & 36.5 \\
\hline
\end{tabular}
\end{table}

\subsection{Wavelet Analysis}

Wavelet analysis is particularly well-suited for studying non-stationary oscillations such as those observed in sunspots \citep{1998BAMS...79...61T}. Unlike Fourier analysis, which assumes stationarity, wavelet analysis provides both temporal and frequency information simultaneously.

For each pixel or region, the light curve was first detrended by subtracting the mean. Wavelet analysis was then performed using the Morlet mother wavelet, which is optimal for detecting oscillatory signals. The wavelet transform of a time series $x(t)$ is defined as:

\begin{equation}
W(s,\tau) = \int_{-\infty}^{\infty} x(t) \psi^*_{s,\tau}(t) dt
\end{equation}

where $\psi_{s,\tau}(t)$ is the Morlet wavelet scaled by $s$ and translated by $\tau$, and $*$ denotes complex conjugation.

The global wavelet spectrum was obtained by averaging the wavelet power over time:

\begin{equation}
\overline{W}^2(s) = \frac{1}{N} \sum_{n=0}^{N-1} |W_n(s)|^2
\end{equation}

From the wavelet power spectrum, two types of periods were extracted:

Dominant period: The period corresponding to the maximum power in the global wavelet spectrum:

\begin{equation}
P_{dom} = \arg\max_{P} \overline{W}^2(P)
\end{equation}

Weighted mean period: The power-weighted average of all periods:

\begin{equation}
P_{wtd} = \frac{\sum_i P_i \cdot \overline{W}^2(P_i)}{\sum_i \overline{W}^2(P_i)}
\end{equation}

The weighted mean period is more physically meaningful because it accounts for the multi-mode nature of solar oscillations, where energy is distributed across multiple periods rather than concentrated at a single frequency.

\subsection{Analysis Methods Comparison}

To assess the robustness of our results, we compared four different analysis methods:

\textbf{Method 1 - Pixel Mean}: Wavelet analysis was performed on each pixel individually, and the resulting periods were averaged across all pixels in each region. This method preserves spatial information and provides a robust estimate of the characteristic period.

\textbf{Method 2 - Pixel Median}: Similar to Method 1, but the median period was used instead of the mean. This method is more robust against outliers caused by noisy pixels or cosmic ray hits.

\textbf{Method 3 - Spatial Mean}: The light curves of all pixels in a region were first averaged spatially, and wavelet analysis was then performed on the average light curve. This method maximizes signal-to-noise ratio but may be affected by phase cancellation if oscillations in different pixels are not coherent.

\textbf{Method 4 - Spatial Median}: The median light curve across all pixels in a region was computed first, followed by wavelet analysis. This method combines the noise robustness of the median with the signal enhancement of spatial integration.

The comparison of these four methods allows us to assess the robustness of measured periods and identify potential systematic biases introduced by different analysis choices.

\section{Results}
\label{sec:results}

\subsection{Period Analysis Results}

\begin{figure}[htbp]
\centering
\includegraphics[width=0.95\textwidth]{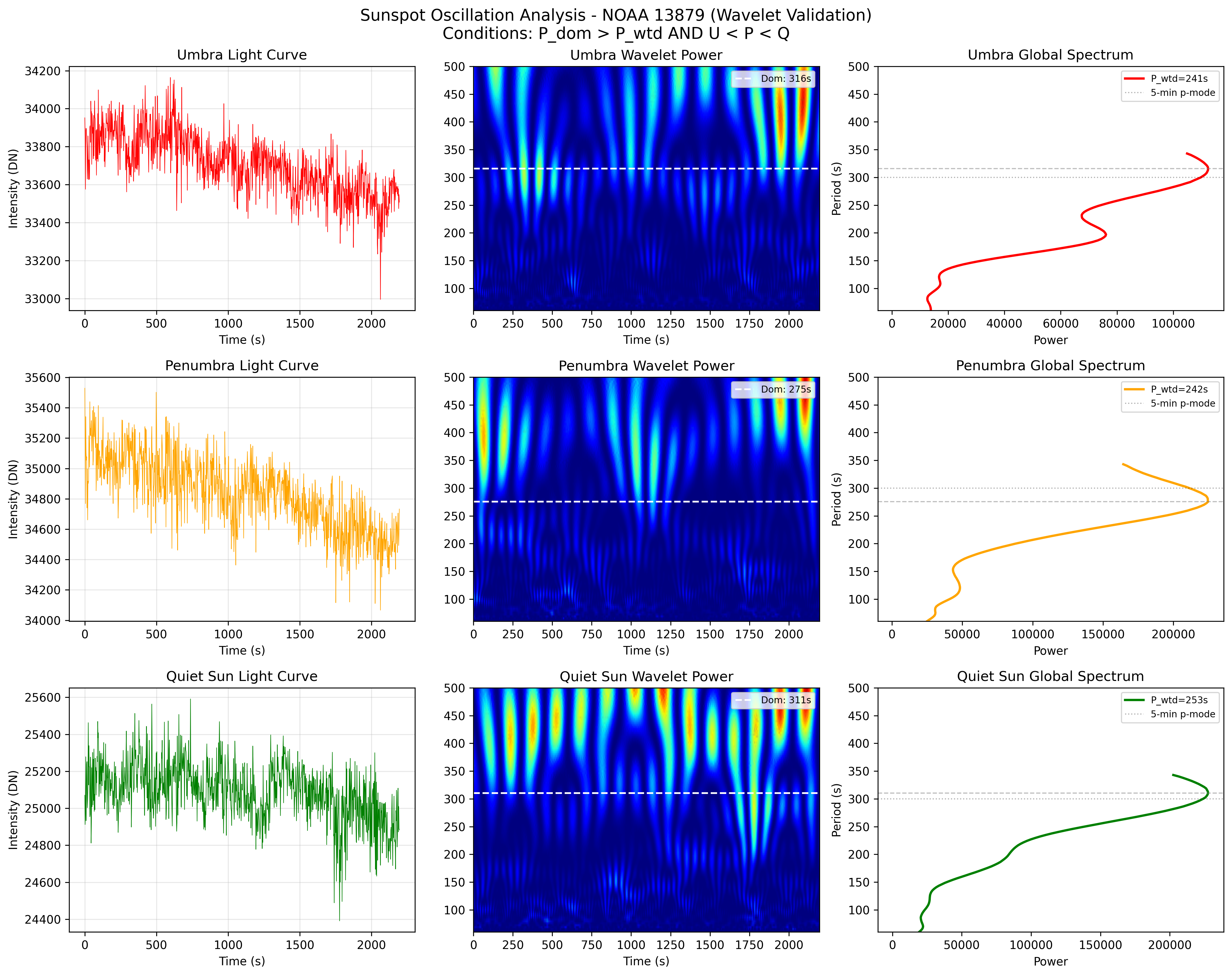}
\caption{Wavelet analysis example for a single pixel in NOAA AR 13837. Left: Light curves of the umbra, penumbra, and quiet Sun. Middle: Wavelet power spectra (Morlet wavelet) as functions of time (x-axis) and period (y-axis). Dashed horizontal lines indicate the 5-minute period, and the legend shows the dominant period. Power is color-coded from blue (low) to red (high). Right: Global wavelet spectra with the weighted mean period ($P_{\rm wtd}$) marked by dashed horizontal lines. The dotted horizontal line indicates the 5-minute (300~s) period.}
\label{fig:wavelet_validation}
\end{figure}

Figure~\ref{fig:wavelet_validation} shows an example of wavelet analysis for a randomly sampled pixel from the umbra, penumbra, and quiet Sun regions of NOAA AR 13837. It is evident that all three regions exhibit clear oscillations, demonstrating that the 8-10~$\mu$m continuum is sensitive to oscillatory signals across different magnetic environments. In this example, the dominant periods are longer than the weighted mean periods. The weighted mean periods for this active region are 241~s (umbra), 242~s (penumbra), and 253~s (quiet Sun). All oscillation period analyses presented in this paper are based on this wavelet analysis methodology.
Having validated the wavelet-based approach with a single-pixel example, we now extend the analysis to all pixels within each region and across all six datasets.

\begin{figure}[htbp]
\centering
\includegraphics[width=0.95\textwidth]{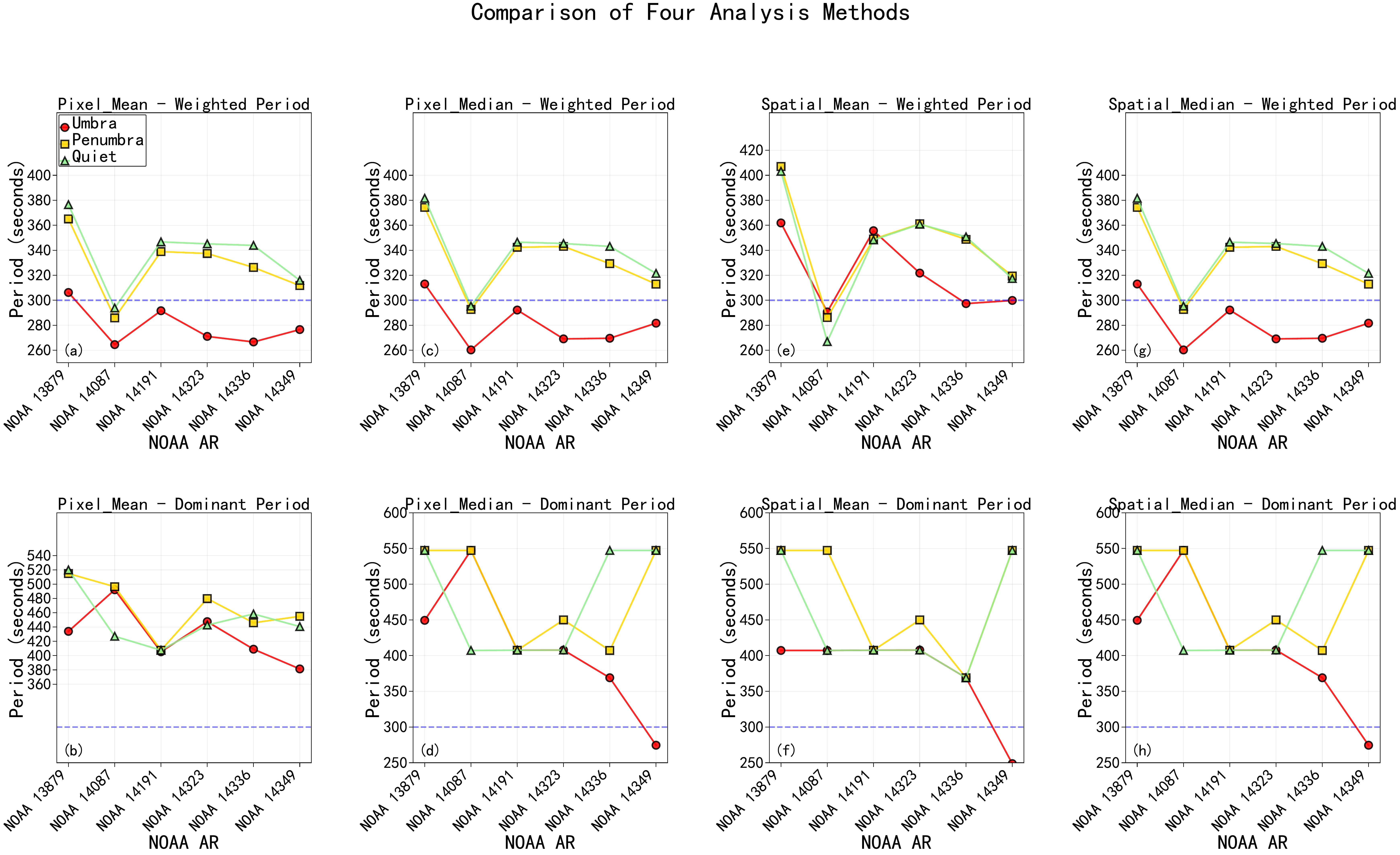}
\caption{Comparison of weighted mean periods (\textbf{top panel}) and dominant periods (\textbf{bottom panel}) across all six datasets using four analysis methods (Pixel Mean, Pixel Median, Spatial Mean, and Spatial Median). \textbf{Top panel:} Weighted mean periods consistently recover the period hierarchy (umbra $<$ penumbra $<$ quiet Sun) in Pixel Mean, Pixel Median, and Spatial Median methods (6/6 datasets, 100\% consistency). The Spatial Mean method shows anomalies in 2/6 datasets where the umbra period exceeds both penumbra and quiet Sun periods, indicating phase cancellation effects. \textbf{Bottom panel:} Dominant periods show large scatter and systematic overestimation (60--230~s longer than weighted mean periods) due to sensitivity to long-period noise and trend components. The gray dashed line at 300~s indicates the theoretical 5-minute p-mode oscillation period.}
\label{fig:period_comparison}
\end{figure}

Figure~\ref{fig:period_comparison} compares the weighted mean periods (top panel) and dominant periods (bottom panel) across all six datasets using four analysis methods. The weighted mean periods are consistently closer to the theoretical 5-minute (300~s) oscillation period than the dominant periods, which tend to be biased toward longer periods. Compared to the weighted mean periods, the dominant periods show systematic overestimation by 60--230~s and exhibit poorer regularity and stability across different datasets and analysis methods, indicating that they are more susceptible to contamination by long-period noise and trend components. These results confirm that the weighted mean period is more physically meaningful and should be used as the primary analysis metric. Therefore, all subsequent analyses are based on weighted mean periods. Nevertheless, it is worth noting that the dominant periods, despite their limitations, consistently show the same trend of shorter periods in the umbra compared to the penumbra and quiet Sun, further supporting the robustness of the observed period hierarchy.

For the weighted mean periods, the period hierarchy depends on the analysis method. The Pixel Mean, Pixel Median, and Spatial Median methods all recover a consistent pattern (umbra (U) $<$ penumbra (P) $<$ quiet Sun (Q)) across all six datasets, with umbra periods ranging from 260--313~s, penumbra periods from 286--374~s, and quiet Sun periods from 294--382~s. However, the Spatial Mean method fails to recover this hierarchy in 2/6 datasets (NOAA 14087 and 14191), where the umbral period exceeds both penumbral and quiet Sun periods due to phase cancellation effects. An additional 3/6 datasets (NOAA 13879, 14323, and 14349) show partial consistency with Spatial Mean (U $<$ P and U $<$ Q satisfied, but P $\ge$ Q). Given that the intrinsic period difference between the penumbra and quiet Sun is very small (typically $<$10~s), such a reversal (P $\ge$ Q) can easily arise from computational uncertainties (e.g., wavelet spectral resolution, finite time series length, or residual noise) and does not undermine the key physical result that the umbra has the shortest period. This method-dependent behavior highlights the importance of choosing an appropriate analysis technique.

Table~\ref{tab:methods_comparison} presents the detailed weighted mean period values corresponding to Figure~\ref{fig:period_comparison} (top panel). The table provides a quantitative summary of the period hierarchy across all four methods and six datasets.

\begin{table}[htbp]
\centering
\caption{Weighted mean periods (in seconds) for all datasets using four analysis methods. Values are shown as Umbra/Penumbra/Quiet. Symbols indicate: $\checkmark$ = full consistency (U $<$ P $<$ Q), $\checkmark_{\rm p}$ = partial consistency (U $<$ P and U $<$ Q, but P $\ge$ Q), $\times$ = inconsistency (U $>$ P or U $>$ Q).}
\label{tab:methods_comparison}
\begin{tabular}{lccccc}
\hline
\textbf{NOAA AR} & \textbf{Pixel Mean} & \textbf{Pixel Median} & \textbf{Spatial Mean} & \textbf{Spatial Median} \\
\hline
13879 & 306/365/376 $\checkmark$ & 313/374/382 $\checkmark$ & 362/407/403 $\checkmark_{\rm p}$ & 313/374/382 $\checkmark$ \\
14087 & 265/286/294 $\checkmark$ & 260/293/295 $\checkmark$ & 291/286/267 $\times$ & 260/293/295 $\checkmark$ \\
14191 & 292/339/347 $\checkmark$ & 292/342/346 $\checkmark$ & 356/349/348 $\times$ & 292/342/346 $\checkmark$ \\
14323 & 271/337/345 $\checkmark$ & 269/343/345 $\checkmark$ & 322/361/361 $\checkmark_{\rm p}$ & 269/343/345 $\checkmark$ \\
14336 & 267/326/344 $\checkmark$ & 270/329/343 $\checkmark$ & 297/349/351 $\checkmark$ & 270/329/343 $\checkmark$ \\
14349 & 277/312/316 $\checkmark$ & 282/313/321 $\checkmark$ & 300/319/317 $\checkmark_{\rm p}$ & 282/313/321 $\checkmark$ \\
\hline
\textbf{Consistency} & 6/6 (100\%) & 6/6 (100\%) & 4/6 (67\%$^{\rm a}$) & 6/6 (100\%) \\
\hline
\multicolumn{5}{l}{$^{\rm a}$Spatial Mean: 1 full ($\checkmark$) [14336] + 3 partial ($\checkmark_{\rm p}$) [13879, 14323, 14349] + 2 inconsistent ($\times$) [14087, 14191]} \\
\end{tabular}
\end{table}

It is worth noting that while the U $<$ P $<$ Q hierarchy is fully satisfied in Pixel Mean, Pixel Median, and Spatial Median methods, the relationship between penumbra and quiet Sun (P vs. Q) is more delicate. As shown in Table~\ref{tab:methods_comparison}, the period difference between penumbra and quiet Sun is intrinsically small, typically less than 10~s. In several cases (e.g., NOAA 13879, 14323, 14349 with the Spatial Mean method), we observe P $\ge$ Q instead of P $<$ Q. Given the small absolute difference, such a reversal can easily arise from computational uncertainties, including wavelet spectral resolution, finite time series length, and residual noise. In fact, if we relax the criterion to only require U $<$ P and U $<$ Q (i.e., the umbra has the shortest period), the consistency improves significantly. Under this relaxed criterion, Pixel Mean, Pixel Median, and Spatial Median still achieve 6/6 (100\%), and even the Spatial Mean method achieves 5/6 (83\%) consistency (full + partial cases). This strongly supports the physical conclusion that stronger magnetic fields (umbra) lead to shorter oscillation periods, regardless of the subtle P--Q ordering.

The Pixel Mean and Pixel Median methods produce remarkably similar results (differences $<$ 5\%), indicating that outliers do not significantly affect the statistical properties of the oscillation periods. The Spatial Median method yields identical results to Pixel Median, suggesting that the median operation commutes with wavelet analysis for sufficiently coherent oscillation signals.

\subsection{Spatial Smoothing Effect}

\begin{figure}[htbp]
\centering
\includegraphics[width=0.9\textwidth]{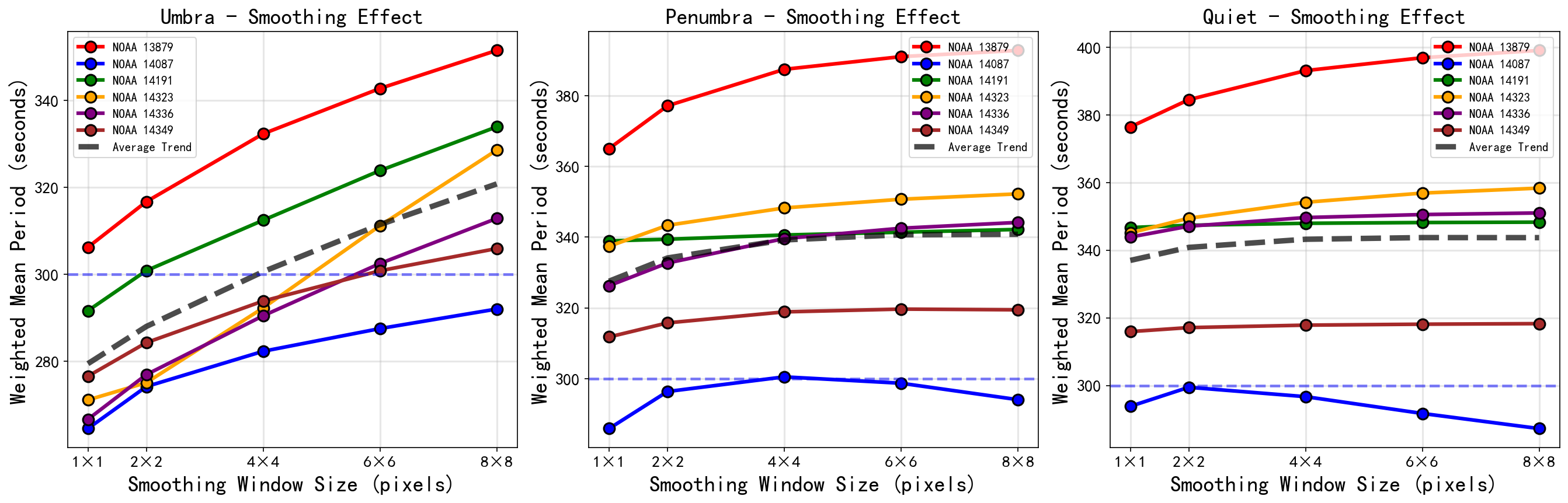}
\caption{Effect of spatial smoothing on weighted mean periods using the Pixel Mean method as an example. The left, middle, and right panels show the results for the umbra, penumbra, and quiet Sun, respectively. The horizontal axis indicates the smoothing window size ($1\times1$, $2\times2$, $4\times4$, $6\times6$, and $8\times8$ pixels). Each panel contains six curves corresponding to the six datasets. The umbra exhibits the strongest sensitivity to smoothing.}
\label{fig:smoothing}
\end{figure}

Figure~\ref{fig:smoothing} shows the effect of spatial smoothing on the measured weighted mean periods using the Pixel Mean method as an example. The other three analysis methods yield qualitatively similar results and are therefore not shown for brevity. 
The figure presents results for all six datasets, organized into three panels: the left panel shows the umbra results, the middle panel shows the penumbra results, and the right panel shows the quiet Sun results. Within each panel, the horizontal axis represents increasing smoothing window sizes ($1\times1$, $2\times2$, $4\times4$, $6\times6$, and $8\times8$ pixels), and the vertical axis shows the measured weighted mean period in seconds. Each panel contains six curves, one for each dataset.

On the whole, as the smoothing window size increases, the measured periods increase monotonically for all three regions (umbra, penumbra, and quiet Sun). This systematic increase is observed consistently across all six datasets.

The umbra shows the highest sensitivity to spatial smoothing, with period increases ranging from +10.4\% to +21.2\% (average +14.8\%) when comparing $1\times1$ to $8\times8$ smoothing. The penumbra shows moderate sensitivity (+1.0\% to +7.6\%, average +4.0\%), while the quiet Sun is least affected ($-2.2\%$ to +6.0\%, average +1.8\%). This differential sensitivity suggests that the umbra contains more small-scale spatial variations in oscillation phase and period that are averaged out by smoothing.

The smoothing effect has important implications for comparing observations from instruments with different spatial resolutions. Studies using lower-resolution data may systematically overestimate oscillation periods, particularly in the umbra. Our results suggest that period measurements should be corrected for resolution effects when comparing across different instruments or when comparing observations with numerical simulations.

\subsection{Peak Ratio and Multi-Mode Oscillations}

The peak ratio, defined as the ratio of the power at the weighted mean period to the total power in the global wavelet spectrum, quantifies the degree to which oscillations are dominated by a single periodic component:

\begin{equation}
R_{\rm peak} = \frac{\overline{W}^2(P_{\rm wtd})}{\sum_i \overline{W}^2(P_i)}
\end{equation}

where $P_{\rm wtd}$ is the weighted mean period and the denominator represents the total spectral power. We use the weighted mean period rather than the dominant period because it is more stable and less susceptible to noise, as demonstrated in Section \ref{sec:results}. Values of $R_{\rm peak} < 0.3$ indicate that no single period dominates, characteristic of multi-mode oscillations where energy is distributed across multiple frequency components \citep{1998BAMS...79...61T}.

\begin{figure}[htbp]
\centering
\includegraphics[width=0.9\textwidth]{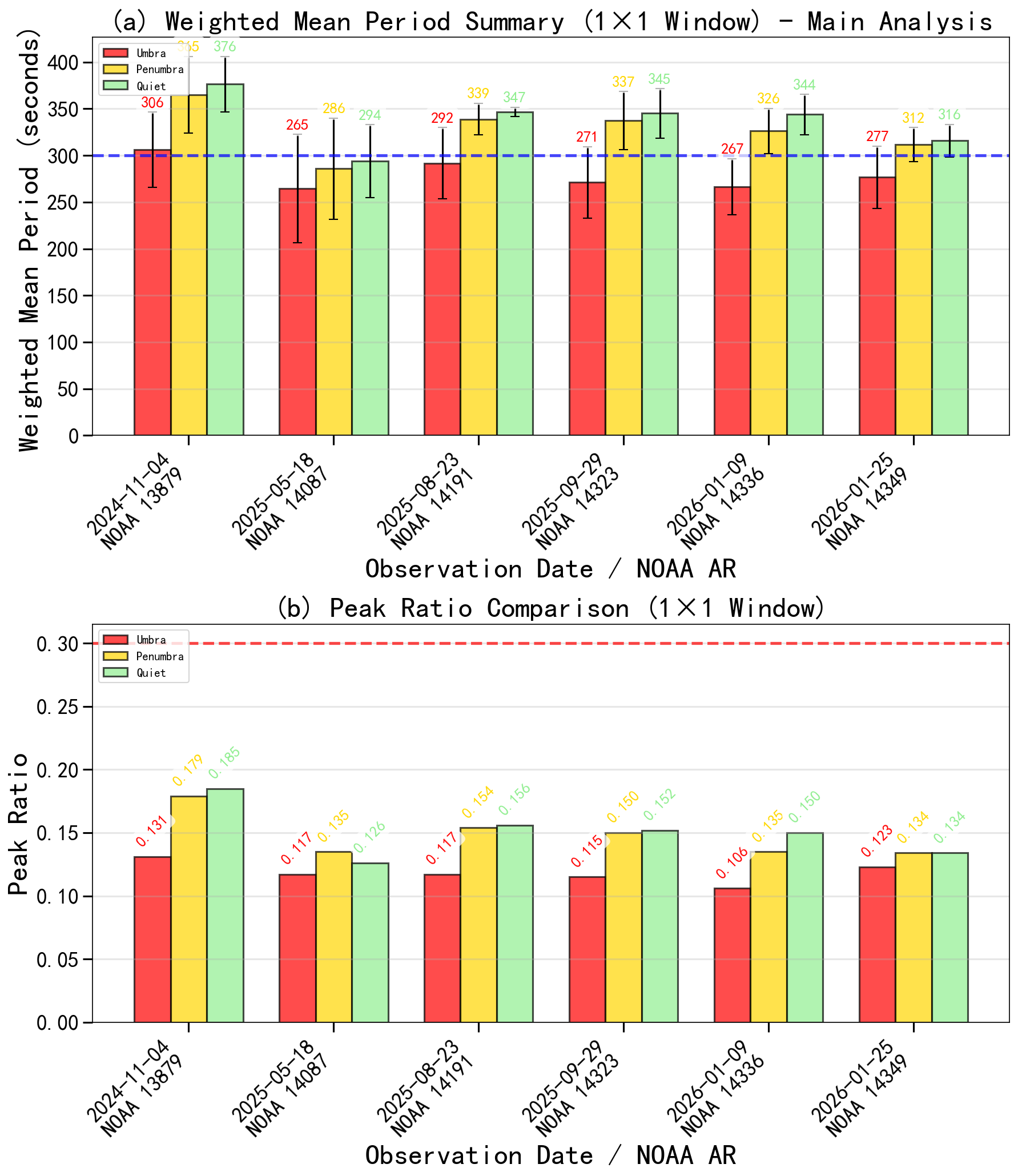}
\caption{Weighted mean periods (top panel) and peak ratios (bottom panel) for all six datasets using the Pixel Mean method as an example. In the top panel, red, yellow, and green symbols represent the umbra, penumbra, and quiet Sun, respectively; the dashed horizontal line at 300~s marks the theoretical 5-minute p-mode oscillation period. In the bottom panel, the same color coding applies, and the dashed horizontal line at 0.3 indicates the threshold for multi-mode oscillations. All measured peak ratios fall below 0.3, with a mean of $0.14 \pm 0.03$. The other three analysis methods yield nearly identical results and are omitted for clarity.}
\label{fig:peak_ratio}
\end{figure}

Figure~\ref{fig:peak_ratio} presents the weighted mean periods (top panel) and peak ratio distribution (bottom panel) across all six datasets and all three regions using the Pixel Mean method as an example. (The other three analysis methods yield nearly identical results and are omitted for clarity.)  The key findings are:

\begin{enumerate}
\item \textbf{All peak ratios are below 0.3}: The maximum observed peak ratio is 0.185 (quiet Sun in NOAA 13879), well below the 0.3 threshold. The mean peak ratio across all measurements is $0.14 \pm 0.03$, confirming that sunspot oscillations are inherently multi-mode in the 8-10~$\mu$m band.

\item \textbf{Regional variation}: The umbra shows lower peak ratios ($0.12 \pm 0.01$) compared to the penumbra ($0.15 \pm 0.02$) and quiet Sun ($0.15 \pm 0.02$), suggesting that umbral oscillations have a broader frequency spectrum, possibly due to multiple wave modes coexisting in the strongly magnetized plasma.

\item \textbf{Implications for analysis}: The low peak ratios justify our choice of using the weighted mean period rather than the dominant period for physical interpretation. In a multi-mode oscillation system, the weighted mean better represents the characteristic oscillation timescale by accounting for contributions from all significant modes.
\end{enumerate}

\section{Discussion}
\label{sec:disc}

\subsection{Physical Interpretation of the Period Hierarchy}

The consistent pattern of umbra period $<$ penumbra period $<$ quiet Sun period observed across all six datasets and three analysis methods (Pixel Mean, Pixel Median, Spatial Median) provides observational evidence for the influence of magnetic fields on oscillation periods in the 8-10 $\mu$m band. This period hierarchy can be understood through the framework of magneto-convection and MHD wave propagation.

In the umbra, where the magnetic field is strongest (typically 2000--3000 G) and nearly vertical, convective motions are strongly suppressed. Oscillations in this region are primarily driven by upwardly propagating slow magneto-acoustic waves guided along the magnetic field lines \citep{2006ApJ...640.1153C, 2017ApJ...836...18C}. The magnetic tension force provides an additional restoring force beyond gas pressure, leading to shorter oscillation periods compared to the quiet Sun. Based on the three methods that fully recover the U $<$ P $<$ Q hierarchy (Pixel Mean, Pixel Median, and Spatial Median; the Spatial Mean method is excluded due to phase cancellation artifacts), our measured umbral periods (260--313 s, or 4.3--5.2 minutes) are systematically shorter than the quiet Sun periods (294--382 s) in the same datasets, confirming the magnetic field's suppressing effect on convective motions.

Compared to the classic three-minute umbral oscillations (periods typically 150--200~s) first discovered in chromospheric observations \citep{1969SoPh....7..351B} and subsequently confirmed by numerous studies (see, e.g., the review by \citealt{2000SoPh..192..373B}), our photospheric measurements at 8-10 $\mu$m show longer periods. This height dependence is expected from the theory of upwardly propagating magneto-acoustic waves: as waves propagate into higher, lower-density atmospheric layers, the acoustic cutoff frequency increase, allowing shorter period waves to propagate in the chromosphere while longer period modes dominate in the photosphere.

In the penumbra, the magnetic field is weaker (1000-1500 G) and more inclined (40-70$^\circ$ from vertical), allowing some convective motions to persist through a filamentary structure of interleaved hot and cold plasma flows \citep{2015ASSL..417.....R}. This results in intermediate periods (286-374 s: based on the Pixel Mean, Pixel Median, and Spatial Median methods, the same as 260-313 s for umbra and 294-382 s for quiet Sun), which reflect a mixture of magneto-acoustic waves and residual convective oscillations. The larger scatter in penumbral periods compared to umbral periods may reflect the intrinsic structural complexity of penumbral filaments.

In the quiet Sun, where there is no strong organized magnetic field, the oscillation spectrum is dominated by the well-known 5-minute p-modes. However, our weighted mean periods for the quiet Sun (294-382 s) are systematically longer than the canonical 300 s value. This can be attributed to two factors: (1) the multi-mode nature of solar oscillations, where the weighted mean includes contributions from longer-period modes, and (2) the higher formation height of the 8-10 $\mu$m continuum, where the acoustic cutoff period is longer due to the lower temperature gradient.

The period difference between umbra and quiet Sun ($\Delta P_{\rm U-Q} = 29-77$ s in our sample) provides a quantitative measure of the magnetic field's influence on oscillation properties. This value may serve as a diagnostic for the magnetic field strength when combined with atmospheric models.

\subsection{Multi-Mode Nature of Sunspot Oscillations}

The consistently low peak ratios observed in this study (all below 0.3, mean $0.14 \pm 0.03$) provide direct evidence that sunspot oscillations in the 8-10~$\mu$m band are inherently multi-mode. Unlike idealised single-mode oscillations often assumed in simplified models, the solar atmosphere supports a superposition of multiple magneto-acoustic and convective modes simultaneously. This multi-mode character has important implications for oscillation analysis.

First, the presence of multiple modes justifies our use of the weighted mean period rather than the dominant period as the primary analysis metric. The dominant period, which simply selects the frequency with maximum power, is sensitive to noise and long-period trends and can be misleading when power is distributed across a broad frequency range. In contrast, the weighted mean period accounts for contributions from all significant modes, providing a more robust and physically meaningful characteristic timescale.

Second, the regional variation in peak ratios-- lower in the umbra ($0.12 \pm 0.01$) compared to the penumbra ($0.15 \pm 0.02$) and quiet Sun ($0.15 \pm 0.02$)----uggests that stronger magnetic fields broaden the oscillation spectrum. This may be due to the coexistence of multiple wave modes (e.g., slow and fast magneto-acoustic waves) in the strongly magnetized umbral plasma, or to enhanced mode coupling in regions with steep magnetic field gradients.

These findings are consistent with previous studies that have demonstrated the multi-mode nature of solar oscillations using different observational diagnostics {\citep{1998BAMS...79...61T}}. Our results extend this understanding to the 8-10~$\mu$m band, confirming that multi-mode behaviour persists throughout the photosphere and into the upper photospheric layers.

\subsection{Limitations}

Several limitations of this study should be acknowledged. First, the temporal coverage of each dataset is approximately 36 minutes, which limits the detection of oscillations with periods longer than about 10 minutes. Second, the spatial resolution (1.5 arcsec pixel$^{-1}$) is moderate compared to other solar instruments, which may affect the identification of fine-scale structures in the penumbra. Third, the wavelet analysis method, while well-suited for non-stationary signals, requires careful interpretation of edge effects near the beginning and end of the time series (the cone of influence). Finally, our analysis is limited to six active regions observed during a specific period (2024-2026); a larger sample spanning different phases of the solar cycle would help generalize the findings.

\section{Conclusions}
\label{sec:concl}

Based on the analysis of six sunspot active regions observed in the 8-10 $\mu$m infrared band with AIMS using wavelet transform methods, we draw the following conclusions.

\begin{enumerate}
\item Period hierarchy: All six datasets show a consistent period sequence: umbra period $<$ penumbra period $<$ quiet Sun period. The measured ranges are:
   \begin{itemize}
   \item Umbra: 260--313~s (4.3--5.2 minutes)
   \item Penumbra: 286--374~s (4.8--6.2 minutes)
   \item Quiet Sun: 294--382~s (4.9--6.4 minutes)
   \item Umbra-Quiet difference: 29--77~s
   \end{itemize}
   This pattern is recovered with 100\% consistency by Pixel Mean, Pixel Median, and Spatial Median methods.

\item The period hierarchy reflects the magnetic field structure: strong vertical fields in the umbra support short-period magneto-acoustic waves, while weaker inclined fields in the penumbra and absent fields in the quiet Sun allow progressively longer period oscillations. The umbra-Quiet Sun period difference ($\Delta P_{\rm U-Q}$) may serve as a diagnostic for magnetic field strength.

\item The weighted mean period is more physically meaningful than the dominant period for characterizing sunspot oscillations, as it accounts for the multi-mode nature of these oscillations. All peak ratios in our study are below 0.3 (mean $0.14 \pm 0.03$), confirming that solar oscillations are inherently multi-mode.

\item Spatial smoothing systematically increases measured periods, with the largest effect in the umbra (10--21\%). The penumbra shows moderate sensitivity (1.0--7.6\%), while the quiet Sun is least affected ($-2.2\%$ to $+6.0\%$). This has important implications for comparing observations from instruments with different spatial resolutions.

\item To our knowledge, this is the first systematic study of sunspot oscillations in the 8-10 $\mu$m mid-infrared band. The measured periods are consistent with upper photospheric formation heights, being systematically longer than chromospheric measurements but consistent with photospheric observations at other wavelengths.
\end{enumerate}

\begin{acknowledgments}
This work was supported by National Key R\&D Program of China 2021YFA1600500, 2022YFF0503001 and 2022YFF0503800, the Strategic Priority Research Program on Space Science, the Chinese Academy of Sciences (Grant No. XDA15320301, XDA15320302, XDA15052200), Natural Science Foundation of China (Grant No 11703042, 11973056, 12003051, 12173049, U1731241, 11427901, 11473039 and 11178016).

We sincerely thank the staff of Huairou Solar Observing Station, including
the engineers, observers, and data processing personnel, for their dedicated
work under the extremely harsh conditions of high altitude and thin air
at the observing site. Their efforts made it possible to obtain the high-quality
scientific data used in this study. We also thank the anonymous referee
for the constructive comments and suggestions that significantly improved
this manuscript.
\end{acknowledgments}

\begin{contribution}
Suo Liu prepared the main manuscript and was responsible for data analysis.
\end{contribution}

\section*{Data Availability Statement}
The data used in this study are available from the Huairou Solar Observing Station data archive at \url{https://sun10.bao.ac.cn/hsos_data/AIMS}.

\bibliographystyle{aasjournalv7}
\bibliography{sunspot_oscillation}
\end{document}